\documentclass{iaus}
\usepackage{graphicx}

\title[TNOs by Stellar Occultation] 
{Search for Small Trans-Neptunian Objects by the TAOS Project }

\author[Chen et al.]   
{
W. P. Chen$^1$,
C. Alcock$^2$,
T. Axelrod$^3$, 
F. B. Bianco$^{2,6}$,  
Y. I. Byun$^4$,
Y. H. Chang$^1$, 
K. H. Cook$^5$, 
R. Dave$^6$, 
J. Giammarco$^6$,
D. W. Kim$^4$, 
S. K. King$^7$, 
T. Lee$^7$, 
M. Lehner$^2$,
C. C. Lin$^1$, 
H. C. Lin$^1$,  
J. J. Lissauer$^8$, 
S. Marshall$^9$, 
N. Meinshausen$^{10}$
S. Mondal$^1$, 
I. de Pater$^{10}$, 
R. Porrata$^{10}$, 
J. Rice$^{10}$, 
M. E. Schwamb$^{6,11}$,
A. Wang$^7$, 
S. Y. Wang$^7$, 
C. Y. Wen$^7$,
Z.W. Zhang$^1$
}

\affiliation{
$^1$Institute of Astronomy, National Central University, Taiwan, 
{\it email: wchen@astro.ncu.edu.tw}  \\[\affilskip]
$^2$Harvard-Smithsonian Center for Astrophysics, USA \\[\affilskip]
$^3$University of Arizona, USA \\[\affilskip]
$^4$Department of Astronomy, Yonsei University, Korea \\[\affilskip]
$^5$IGPP, Lawrence Livermore
National Laboratory, USA \\[\affilskip]
$^6$Department of Physics \& Astronomy,
University of Pennsylvania, USA  \\[\affilskip]
$^7$Institute of Astronomy and Astrophysics, Academia Sinica, Taiwan  \\[\affilskip]
$^8$NASA Ames Research Center, USA  \\[\affilskip]
$^9$Stanford Linear Accelerator Center, USA  \\[\affilskip]
$^{10}$University of California, Berkeley, USA \\ [\affilskip]  
$^{11}$California Institute of Technology, Pasadena, CA
}

\pubyear{2007}
\volume{236}  
\pagerange{xxx}
\date{?? and in revised form ??}
\jname{Near-Earth Objects, Our Celestial Neighbors:~Opportunity and Risk}
\editors{Andrea Milani, Giovanni B. Valsecchi, \& David Vokrouhlicky, eds.}
\begin{document}

\maketitle

\begin{abstract}
The Taiwan-America Occultation Survey (TAOS) aims to determine the 
number of small icy bodies in the outer reach of the Solar System by 
means of stellar occultation.  An array of 4 robotic small 
(D=0.5 m), wide-field (f/1.9) telescopes have been installed at Lulin Observatory in Taiwan 
to simultaneously monitor some thousand of stars for such rare occultation events.  
Because a typical occultation event by a TNO a few km across will last for only a fraction of 
a second, fast photometry is necessary.  A special CCD readout scheme has been devised to allow 
for stellar photometry taken a few times per second.  Effective analysis pipelines 
have been developed to process stellar light curves and to correlate any possible flux 
changes among all telescopes.  A few billion photometric measurements have been 
collected since the routine survey began in early 2005.  Our preliminary result of 
a very low detection rate suggests a deficit of small TNOs 
down to a few km size, consistent with the extrapolation of some recent studies of 
larger (30--100~km) TNOs.     
\end{abstract}

\firstsection 
\section{The Taiwan-America Occultation Survey (TAOS) Project}

As of September 2006, there are over 1000 Trans-Neptunian Objects (TNOs) known 
(www.boulder.swri.edu/ekonews).  Because of their large distances and small sizes, 
only the brightest, hence the largest, TNOs could be detected by their reflected sunlight.  
The faintest objects detected so far correspond to a size of about a few tens of km.  
There seems a deficit of TNOs in this size range in comparison with larger ones, 
perhaps as a result of collisional disruption (\cite[Bernstein et al. 2004]{ber04}). 
To determine the number of even smaller TNOs, TAOS has implemented 
an array of 4 small (aperture 0.5 m), wide-field (f/1.9) telescopes at Lulin Observatory in Taiwan 
to monitor for chance stellar occultation by TNOs (\cite[King et al. 2001]{kin01}, 
\cite[Alcock et al. 2003] {alc03}).  TNOs as small as 1--2~km across should be detectable 
by TAOS.  The occurrence rate of occultation events will provide 
constraints on the number and spatial distribution of TNOs.  

Occultation by a TNO a few km across lasts for a fraction of a second, so rapid photometry is necessary.   
The 4 TAOS telescopes are located with a maximum separation of $\sim 100$~m, too close 
to detect the shadow speed by timing difference at different telescopes, but allowing coincidence 
detection by multiple telescopes to reduce the false alarm rate. 
In addition to the TNO science, the huge TAOS database, some 100 GB per night, should be valuable
in other studies such as stellar variability.  Our robotic system is also responding to 
gamma-ray bursts.

\section{Current Status of the TAOS Project} 
 
\subsection{Data Acquisition and Adaptive Aperture Photometry}

The TAOS CCD camera, instead of reading out the entire chip as in a regular imaging observation, 
continuously integrates (pause) and reads out a block of pixels (shift) one at a time, 
while the shutter remains open and the telescope tracks on the target field.  
This electronic ``pause-and-shift" scheme in effect produces a sequence of snapshots and allows 
for stellar photometry at a rate up to a few hertz (\cite[Chen et al. 2003]{che03}).   
The image thus obtained is crowded because all the stars in the field of view are ``folded" into 
each readout block.  The shift of electrons and the open shutter also cause a star to leave a streak, which 
interferes with the signal of a ``nearby" star that appears to be adjacent in the  
readout block, though the two stars in fact may be widely separated in the sky.  
In addition, photons from either a neighboring star or a patch of sky are recorded at different times.  

\begin{figure}
 \vspace{-3mm}
 \centering
 \includegraphics[width=\textwidth]{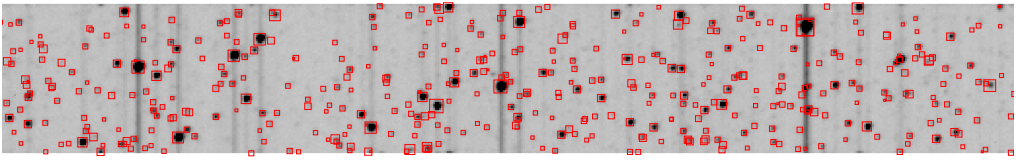}
  \vspace{-5mm}
  \caption{The photometry aperture mask to process the TAOS data.  Each star has a different 
            aperture size for a maximum signal-to-noise light curve.  The position of the aperture
            changes according to the centroid of the star to accommodate image motion. 
    }
  \label{fig:mask}
\end{figure}

A photometry pipeline has been developed to deal with the crowdedness and temporal/spatial 
blending in the data.  For each star, an optimal aperture size is used to minimize
contamination from neighboring stars or streaks.  An appropriate patch of sky is chosen  
for subtraction.  An ``aperture mask" is then applied to measure the fluxes of all the
stars (Fig.~\ref{fig:mask}).  The position of the aperture keeps centered on, i.e., 
being adaptive to, the peak pixel of a star from one readout block to another, so as 
to compensate for image motion.  
This Adaptive Aperture Photometry routine uses square apertures so accuracy is compromised---not 
a serious problem because TAOS images with a $3''$ pixel scale are undersampled---but it  
is very efficient, capable in real time to process about a thousand stars sampled at 5~Hz.  
Other analysis methods are being evaluated.   

\subsection{Event Detection --- The Rank Statistics} 

\begin{figure}
 \centering
 \includegraphics[width=3.8in]{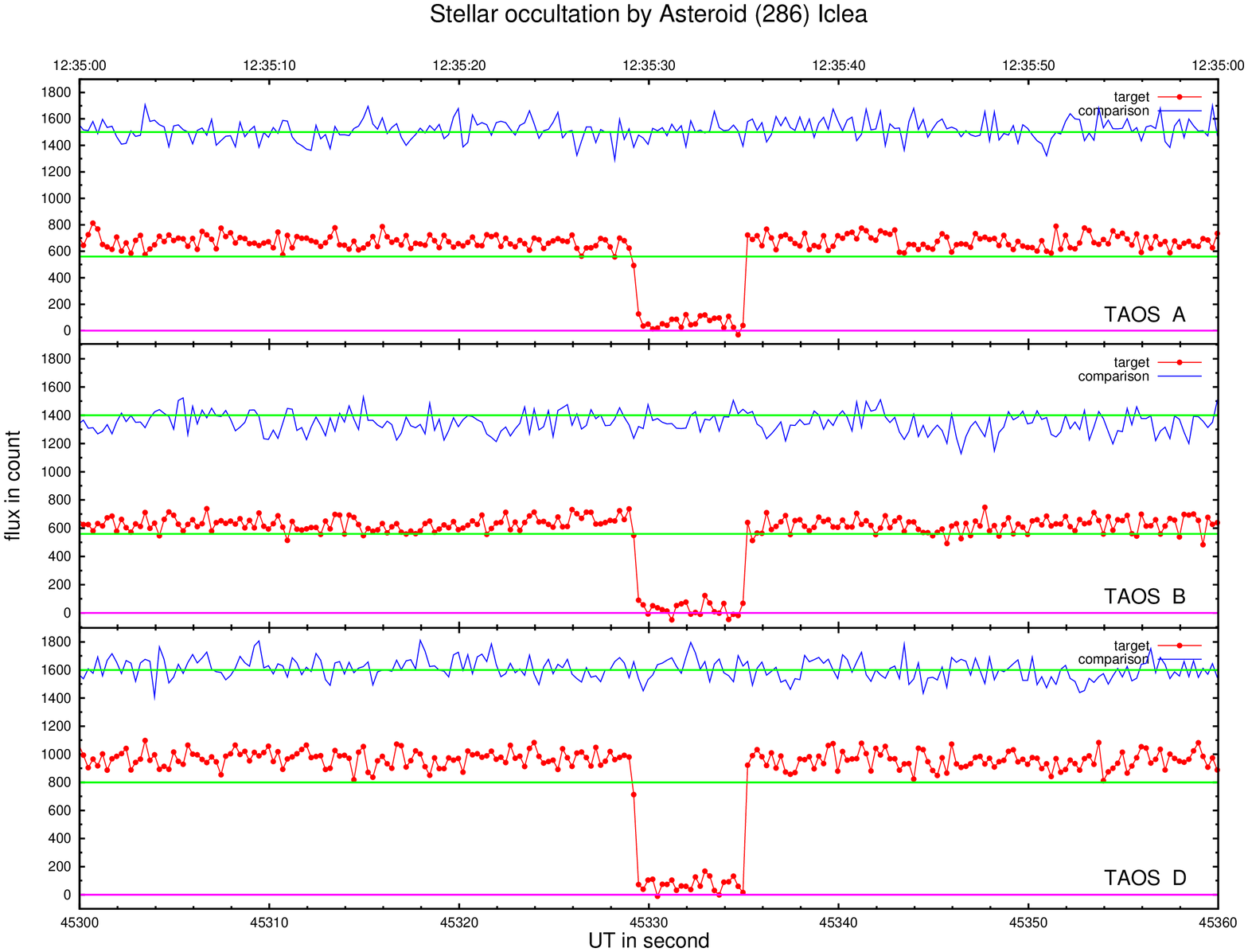}
  \vspace{-5mm}
  \caption{A showcase data of the star TYC\,076200961 (m$_{\rm V} \sim$~11.83~mag) occulted  
                by (286) Iclea (m$_{\rm V} \sim$~14.0~mag at the time) on 6 February 2006 
                observed by 3 TAOS telescopes.  Data were taken at 4~Hz, and the 
                duration of the event was estimated to be $\approx 5.75$~s.    
    }
  \label{fig:iclea}
\end{figure}

Fig.~\ref{fig:iclea} shows the occultation of the star TYC\,076200961
by the asteroid (286) Iclea observed on 
6 February 2006.  TAOS observed this predicted event successfully with 3 telescopes.
Iclea is known to have a diameter of 97~km, and the TAOS system 
detected the event readily.  

For short events, e.g., within one data point, we employ a nonparametric approach to 
identify simultaneous flux drops in stellar light curves.     
The flux at a given time is ranked among all data points.  
The rank statistic is then the product of the ranking orders of all telescopes.  For example, 
if there are $N$ data points from one telescope, the
lowest flux has the rank $R=1$ and the highest
$R=N$.  The rank statistic for data point $i$ is then 
\begin{equation}
  Z_{i} = -\log ( \prod_{k=1}^{N_{\rm tel}} \,  \frac{R_{i}^{k}}{N} ) 
        = \sum_{k=1}^{N_{\rm tel}} ( -\log \, \frac{R_{i}^{k}}{N}), 
\end{equation}
where $N_{\rm tel}$ is the total number of telescopes, and both the 
multiplication $\prod$ and summation $\sum$ are over $N_{\rm tel}$.   The 
quantity $Z$ approximates to a Gamma distribution, if the noise between 
telescopes is independent.    
A probable occultation event would stand out as an ``outlier" 
against an otherwise random distribution.

\subsection{Expected Event Rate}

The detected rate of TNO occultation events depends on the actual occurrence rate
and the detectability.  Relevant parameters include: 
(1)~The surface number density (our goal) and angular size distribution of TNOs.   
(2)~The surface number density and angular size distribution of background stars.
Both these depend on the Galactic line of sight.  With our current instrument 
setup, the 5~Hz observations reach about R=14~mag, which gives within the 3~deg$^{2}$ 
field of view typically several hundred to a few thousand stars in a target 
field.  A dense field would be favorable for occultation but would create images too 
crowded for analysis. The angular size of a star can be estimated, for example, 
by its optical and infrared colors (\cite[van Belle 1999]{van99}).   
Most stars have an angular size less than 0.1 milliarcsecond (mas).  For reference, 
a TNO at 50~AU with an angular size of 0.1 mas has a physical diameter of $\sim 4$~km. 
Items (1) and (2) together specify the probability of area overlap (geometric consideration 
for occultation), plus the diffraction effect (\cite[King et al. 2006]{kin06}; \cite[Lehner et al. 
2006]{leh06}), which tends to smooth out the flux 
drop and becomes important for small or distant TNOs.  
The extrapolated number density of small ($> 1$~km) TNOs varies widely, ranging from 
$10^{2}$ to $10^{6}$ per deg$^2$ (\cite[Bernstein et al. 2004]{ber04}).  
This amounts to about 1 event per few days to 1 event over a few years.  
(3)~The shadow speed.  Observing toward the opposition for example, as opposed to the quadrature, 
maximizes the chance of occultation but the event duration is short so  
difficult to detect.      
(4)~CCD integration time and sampling rate, etc.  These affect the limiting magnitude,  
hence the number of stars observable, and the capability to resolve a short flux drop.    

\begin{figure} [h]
 \centering
  \includegraphics[width=0.34\textwidth,angle=-90]{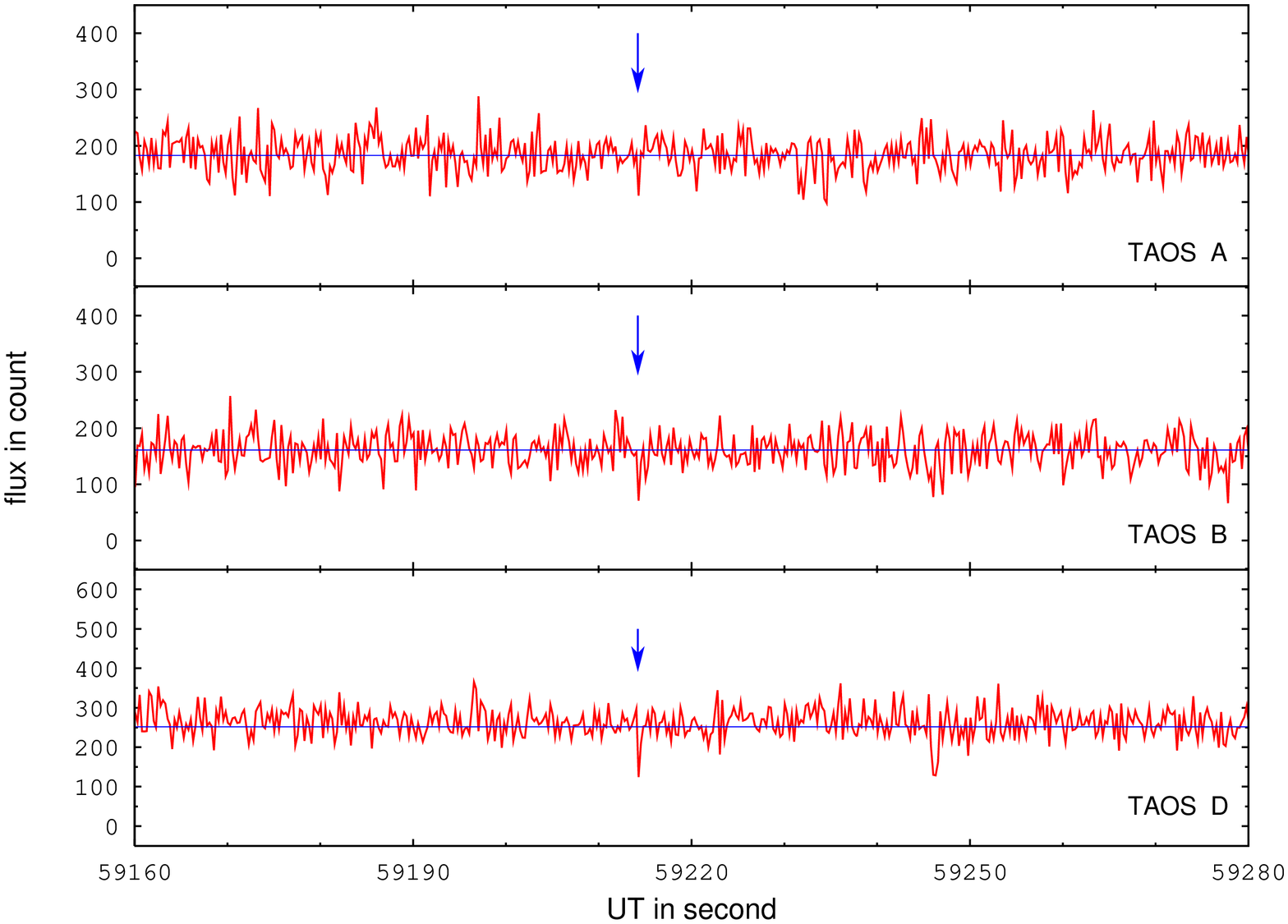}
  \includegraphics[width=0.34\textwidth,angle=-90]{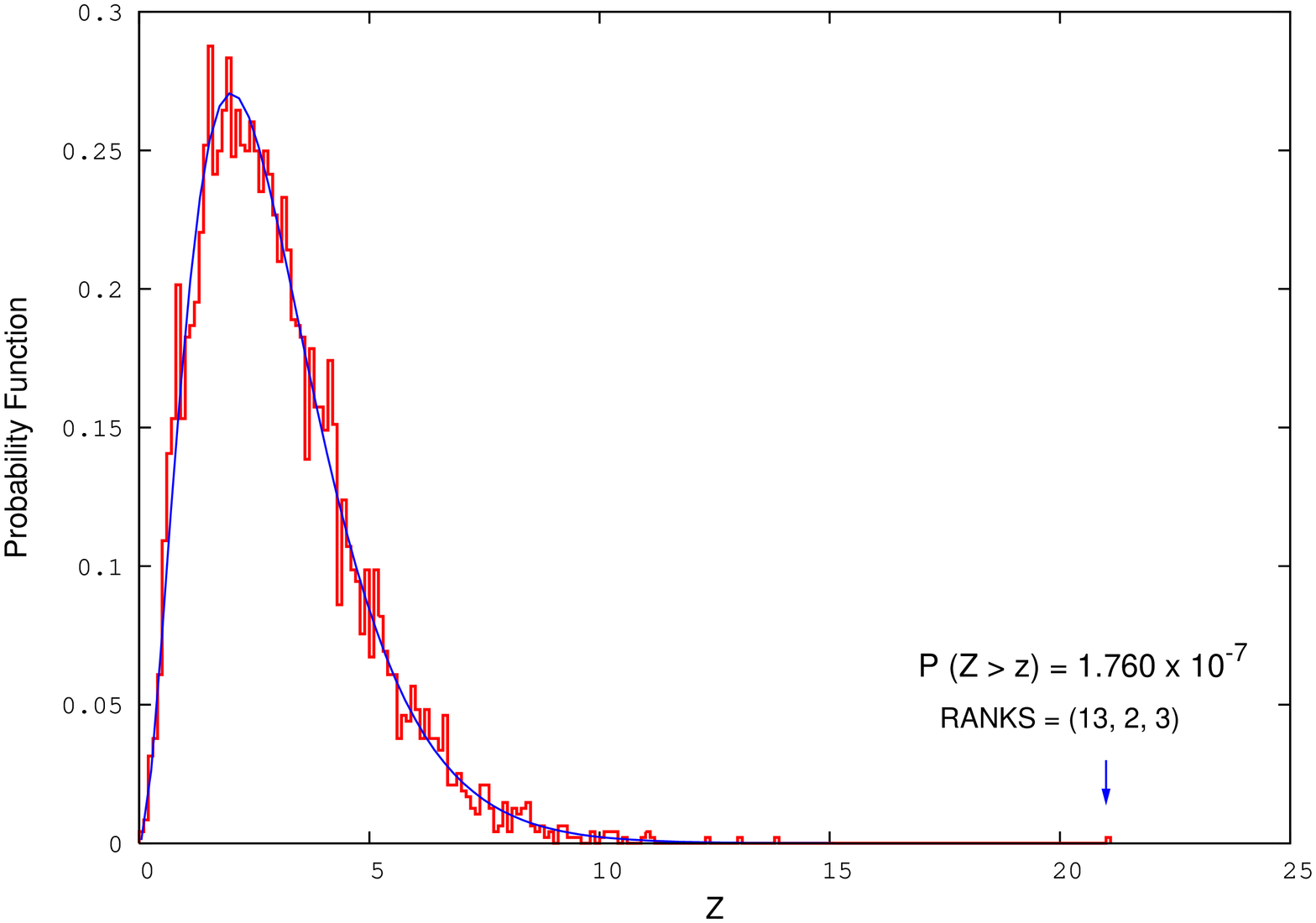}
  \caption{(Left) Light curves of a suspected TNO occultation event observed on 7 April 2005.  
            (Right) Histogram of the rank statistics (2.1).  This event ranks 
             (13, 2, 3), respectively, in 3 telescopes out of 4728 data points 
             sampled every 0.25~s.  The smooth line is the Gamma distribution.  
             The arrow marks the outlier event.  }
  \label{fig:event}
\end{figure}

Since 2005, TAOS with 3 telescopes has collected some 
billions of photometric measurements.  Preliminary analysis shows a very low 
detection rate.  A suspected event is shown 
in Fig.~\ref{fig:event} where a relatively faint star 
(R$\sim $13.5~mag, RA$=13^{\rm h}46^{\rm m}26.7^{\rm s}$ 
and Decl.$=-10^{\circ}50'31''$) was detected to have a 39\%, 56\%, and 50\% flux drop, 
respectively, in 3 telescopes.  The probability is, according to the Gamma approximation, 
$1.7 \times 10^{-7}$ that the coincident flux drops were caused by
chance.  Given our high data rate, however, we cannot be highly confident that this was an actual 
occultation event.  A fourth telescope, which is expected to be in service in early 2007, 
would increase greatly the credibility of a detection. 
Our result suggests a deficit down to 
km-sized TNOs.  It is not clear at the moment how our low detection rate reconciles with the claim of  
numerous even smaller ($< 100$~m) TNOs (\cite[Chang et al. 2006]{cha06}).

\noindent 
The NCU group acknowledges the NSC grant 95-2119-M-008-028.  
KHC's work was performed under the auspices of the US DOE, by the Univ.of
California, LLNL under contract No. W-7405-Eng-48.


\end{document}